\documentclass[aps,prd,twocolumn,showpacs,nofootinbib]{revtex4}
\newcommand{\ar}{\arrowvert}
\newcommand{\ra}{\rangle}
\newcommand{\la}{\langle}
\newcommand{\da}{\dagger}

\newcommand{\cd}{\! \cdot \!}
\newcommand{\be}{\begin{equation}}
\newcommand{\ee}{\end{equation}}
\newcommand{\ba}{\begin{eqnarray}}
\newcommand{\ea}{\end{eqnarray}}

\usepackage{graphicx}
\usepackage{dcolumn}
\usepackage{bm}

\begin{document}
\title{Hyperfine meson splittings:
chiral symmetry versus transverse gluon exchange \\}

\author{Felipe J. Llanes-Estrada } \email{fllanes@fis.ucm.es}
\affiliation{Departamento de F\'{\i}sica Te\'orica I,  Universidad
Complutense, 28040 Madrid, Spain}
\author{Stephen R. Cotanch} \email{cotanch@ncsu.edu}
\affiliation{Department of Physics, North Carolina State University,
Raleigh, NC 27695}
\author{Adam P. Szczepaniak} \email{aszczepa@indiana.edu}
\affiliation{Department of Physics and Nuclear Theory Center, Indiana
University, Bloomington, IN 47405}
\author{Eric S. Swanson} \email{swansone@pitt.edu}
\affiliation{Department of Physics and Astronomy, University of
Pittsburgh, Pittsburgh, PA 15260 \\
 and Jefferson Lab, 12000 Jefferson Ave, Newport News, VA 23606}

\date{\today}

\begin{abstract}
Meson spin splittings are examined within an effective Coulomb gauge QCD Hamiltonian
incorporating chiral symmetry and a transverse hyperfine interaction necessary
for heavy quarks. For light and
heavy quarkonium systems the 
pseudoscalar-vector meson spectrum is generated by approximate BCS-RPA
diagonalizations.  This relativistic formulation
includes both $S$ and $D$ waves for the vector mesons which generates a set of coupled
integral equations.  A smooth transition from  the heavy to the light quark regime  is found
with chiral symmetry dominating the
$\pi$-$\rho$ mass difference.  A reasonable description of the observed meson spin
splittings and chiral quantities, such as the quark condensate and the $\pi$ mass,
is obtained.  Similar comparisons with TDA diagonalizations, which violate chiral symmetry,
are deficient for light pseudoscalar mesons indicating the need
to simultaneously include both chiral symmetry and a hyperfine interaction.
The $\eta_b$ mass is predicted to be around 9400 MeV consistent 
with other theoretical expectations and above the unconfirmed 9300 
MeV candidate. Finally, for comparison with lattice results, the $J$ reliability
parameter is also evaluated.
\end{abstract}


\maketitle

\section{Introduction}

The hyperfine interaction has a long and distinguished history beginning with
the hydrogen atom where it correctly describes the transition responsible
for the famous ``21-centimeter line'' in microwave astronomy. 
Taking the non-relativistic reduction of the one-photon-exchange interaction, 
the hyperfine potential has the form
\begin{equation}
V_{hyp} = {\rm const} \ \frac{\bm{\sigma}_1}{M_1} \cdot \frac{\bm{\sigma}_2}{M_2}\delta(\bm{r})
\label{Vhyp}
\end{equation}
for particles of mass $M_i$ and spin $\bm{\sigma}_i$. 
This potential gives
an accurate description of 
the triplet-singlet splitting in positronium. When implemented in the simple additive quark
model having meson mass $M$ and constituent quark masses ${\cal M}_{q= u, d, s}$ ,
\begin{equation}
M = {\cal M}_q +{\cal M}_{\bar{q}} + A \frac{\bm{\sigma}_q} {{\cal M}_q} \cdot
\frac{\bm{\sigma}_{\bar{q}}} {{\cal M}_{\bar{q}}}
\end{equation}
it produces a remarkably good description of the light meson spectrum using $A \approx 
160 \ {\cal M}^2_u $ MeV. However, to reproduce the splitting in charmonium, using
${\cal M}_{q = c}$ $\approx 1500 $ MeV, requires a
hyperfine strength of at least $4  A$, while a
similar model for baryons uses a much weaker value,  approximately $A/3$~\cite{grif}.
Hence attempts to comprehensively describe spin splittings in hadrons with a simple hyperfine
interaction leads to over an order of magnitude variation in the potential strength.  Although
this may  merely reflect the  simplistic nature of the additive quark model, more extensive models
also have difficulty in obtaining a consistent description for both mesons and baryons with the
same hyperfine interaction. Possibly related, other problems arise when considering the hyperfine
interaction in quark model applications. For example, the non-relativistic reduction of the one
gluon exchange interaction between quark pairs gives rise to  hyperfine, spin-orbit and tensor
interactions at order
$1/{\cal M}_q^2$.  Unfortunately this structure does not describe heavy meson spin
splittings well and must be supplemented with a spin-orbit term which is argued~\cite{gi}
to emerge from the reduction of a  scalar confinement potential. Although this prescription
is reasonably successful for heavy mesons, it generates too large spin-orbit splittings in
baryons~\cite{ci} and the sign of the scalar confinement potential must be
inverted~\cite{sspot}.

Further clouding this matter is the role of chiral symmetry.  Because the pion is regarded
as the Goldstone boson of broken chiral symmetry,  the
$\pi$-$\rho$ mass difference should be dominated by the same nonperturbative dynamics 
producing spontaneous chiral symmetry breaking~\cite{Scadron,brs96,mrt98,csref}.  This observation
is in conflict with the hyperfine splitting in typical constituent (non-chiral) quark models
which  apply  Eq.~(\ref{Vhyp}) to light hadrons.  Indeed in such models the
hyperfine potential  plays a dual role of generating spin splittings
and producing a very light, ``chiral'' pseudoscalar meson.  Because of the quite large, 
over 600 MeV,
$\pi$-$\rho$ and 400 MeV $K$-$K^*$ mass differences, an appreciable hyperfine splitting is
required. This in turn makes it necessary to use a more complex interaction, rather than
a simple potential, to simultaneously describe hadrons not governed by chiral symmetry, such as
excited state light mesons, heavy mesons and baryons which all have smaller spin splittings,
roughly from 50 to 300 MeV. As discussed in Ref.~\cite{ess}, the nonperturbative aspects of spin
splittings in light quark hadrons, especially those involving pseudoscalar mesons, clearly
requires a more sophisticated model treatment.   

The purpose of this paper is to address the above issues and to provide a deeper
understanding of meson spin splittings by examining
the hyperfine interaction in a theoretical framework which
incorporates chiral symmetry.  A related goal is to also provide an
improved hadron approach embodying many of the features of QCD with a
minimal number of parameters (i.e. current quark masses and one or two dynamical
constants). This formulation is based on the Coulomb gauge Hamiltonian of QCD and approximate
diagonalizations using the Bardeen-Cooper-Schrieffer (BCS),
Tamm-Dancoff (TDA) and random phase approximation (RPA) many-body techniques. 
These methods have been previously applied to chiral symmetry breaking~\cite{orsay},
glueballs~\cite{gbs,lcbrs,gss}, hybrids~\cite{sshyb,lcplb}  and mesons~\cite{LC1,LC2,ls2}. 
The renormalization group methodology has also been applied\cite{ssrenorm,rob} to
improve this formalism. Finally,  this work incorporates an effective QCD longitudinal
confining potential~\cite{sspot} along with a
generalized version of the transverse hyperfine interaction employed in an
earlier hyperfine study~\cite{sshyp}.

An  important aspect of the following discussion is that the random phase approximation
is capable of describing chiral symmetry breaking. Indeed, the 
RPA pion mass, $M_\pi$, satisfies the
Gell-Mann-Oakes-Renner relation dictated by chiral symmetry
\begin{equation}
M_\pi =\left[ -\frac{2 m_q \la \bar{q} q \ra}{f_\pi^2} \right]^{1/2} \ ,
\label{gmor}
\end{equation}
where $m_q$ is the current quark mass, $f_\pi$ is the pion decay constant and
$\la \bar{q} q \ra \equiv \la 0 |\bar{\Psi} \Psi | 0\ra$ is the quark
condensate.   In contrast, the TDA 
does not respect chiral symmetry (the RPA, but not TDA, meson field operator 
commutes with the chiral charge~\cite{LC1,LC2}).
Comparing RPA and
TDA masses therefore permits a quantitative assessment of the relative
importance of
both chiral symmetry and the hyperfine interaction for the $\pi$-$\rho$
mass splitting.
A key result  of this work is that chiral symmetry  dominates this
splitting and
accounts for about 400 MeV of the mass difference. Another is that with the same
interaction,  the mass
differences between the radially excited $\pi$ and $\rho$ states are also reproduced,
as well as the pseudoscalar and vector states in charmonium and bottomonium, thereby
demonstrating the universality of this approach. Lastly, the inclusion of the hyperfine
interaction improves the model description of the quark condensate.

In the next section the model
Hamiltonian
is specified and the  BCS and RPA equations are formulated.  Section III
presents numerical results and details sensitivity to different
hyperfine interactions.  Finally, conclusions are presented in section IV.

\section{Effective Hamiltonian and equations of motion}
\subsection{Model Hamiltonian}

As discussed above, the model Hamiltonian is taken to be that of Coulomb gauge
QCD. The Coulomb potential is evaluated self-consistently in the mean-field
gaussian variational ansatz~\cite{sspot}.  The general form of
this effective Hamiltonian in the combined quark and glue sectors is 
\begin{eqnarray}
H_{eff} &=& H_q + H_g +H_{qg} + V_{C}   \\
H_q &=& \int d{\bf x} \Psi^\dagger ({\bf x}) ( -i \bm{\alpha} \cdot \bm{\nabla}
+ m_q \beta ) \Psi ({\bf x})   \\
H_g &=& Tr  \int d {\bf x}\left[ {\bf \Pi}^a({\bf x})\cdot {\bf
\Pi}^a({\bf x}) +{\bf B}^a({\bf x})\cdot{\bf B}^a({\bf x}) \right] \\
H_{qg} &=&  g \int d {\bf x} {\bf J}^a ({\bf x})
\cdot {\bf A}^a({\bf x}) \label{Hqg} \\
V_C &=& -\frac{1}{2} \int d{\bf x} d{\bf y} \rho^a ({\bf x}) \hat{V}(\ar {\bf x}-{\bf y} 
\ar ) \rho^a ({\bf y})   \ .
 \label{model}
\end{eqnarray}
Here $g$ is the QCD coupling, $\Psi$ is the quark field, ${\bf A}^a$ are the 
gluon fields satisfying the transverse gauge condition,
$\mbox{\boldmath$\nabla$\unboldmath}$ $\cdot$
${\bf A}^a = 0$, $a = 1, 2, ... 8$, 
${\bf \Pi}^a $ are the conjugate fields and ${\bf B}^a$ are 
the nonabelian magnetic fields 
\begin{eqnarray}
{\bf B}^a = \nabla \times {\bf A}^a + \frac{1}{2} g f^{abc} {\bf A}^b \times {\bf A}^c \ .
\end{eqnarray}
The color densities, $\rho^a({\bf x})$, and quark color currents, ${\bf J}^a$, are
related to the fields by 
\begin{eqnarray}
\rho^a({\bf x}) &=& \Psi^\dagger({\bf x}) T^a\Psi({\bf x}) +f^{abc}{\bf
A}^b({\bf x})\cdot{\bf \Pi}^c({\bf x}) \\
{\bf J}^a &=& \Psi^\dagger ({\bf x}) \mbox{\boldmath$\alpha$\unboldmath}T^a \Psi ({\bf x})
\ ,
\end{eqnarray}
where
$T^a = \frac{\lambda^a}{2}$ and $f^{abc}$ are  the
$SU_3$ color matrices and structure constants, respectively.

Since this work focuses on the quark sector $H_g$ is omitted and the
quark-glue interaction, $H_{qg}$, is replaced by an effective transverse
hyperfine potential, $V_T$,
discussed below.  The choice for the longitudinal Coulomb potential, $V_C$,
then completes specification of the model. 

To lowest order in $g$, the Coulomb gauge potential $\hat{V}$
is simply proportional to $1/r$. As is well known,
in a few body truncation this is insufficient for confinement and fails to
reproduce the phenomenological meson spectrum. Using the Cornell potential for $\hat V$
resolves these issues but produces an ultraviolet behavior necessitating the
introduction of a model momentum cutoff.
Instead, an improved dynamical treatment~\cite{sspot} is adopted in which both
the gluonic quasiparticle basis and the confining
interaction were determined self-consistently and, through
renormalization, accurately reproduced the lattice Wilson loop potential.
The resulting interaction  has a renormalization
improved
short ranged behavior and  long-ranged confinement.
It is similar to the Cornell potential and has a  numerical representation in momentum
space  that is accurately fit by the analytic form
\begin{equation} 
V(p) = \left\{ \begin{array} {ll} C(p)\equiv -\frac{8.07}{p^2} \, \frac{{\rm log}^{-0.62} \left(
\frac{p^2}{m_g^2}  +0.82 \right) }{{\rm log}^{0.8}\left(
\frac{p^2}{m_g^2} +1.41\right)  } & {\rm for}  \ p >m_g \\
L(p) \equiv -\frac{12.25 \, m_g^{1.93}}{p^{3.93}}  & {\rm for} \ p <m_g \ . \end{array}
\right.
\label{SSpotential} 
\end{equation}
The low momentum component is numerically close to a pure linear potential,
$L \simeq -8 \pi \sigma/p^4 $.  The
other term represents a renormalized high energy Coulomb tail. The only
free parameter
is $m_g \approx 600$ MeV which sets the scale of the theory and 
is equivalent to a string
tension.

Both the exact and model QCD Coulomb gauge Hamiltonians do not explicitly contain a
hyperfine-type interaction, however, perturbatively integrating out gluonic degrees of
freedom generates a quark 
hyperfine interaction with the form $\bm{\alpha}_1 \cdot \bm{\alpha}_2$.  One can also
formally generate this Lorentz structure using Maxwell's equations to substitute for the gluon
fields in Eq. (7).
More generally, in the Hamiltonian formalism an effective hyperfine interaction arises from
nonperturbative mixing of gluonic
excitations (such as hybrids) with the quark Fock space components of a hadron's wavefunction.
However, the $\bm{\alpha}_1\cdot\bm{\alpha}_2$  Lorentz structure is expected to
persist~\cite{aspk} and  Ref.~\cite{aspk} obtains a  $\bm{\alpha}_1 \cdot
\bm{\alpha}_2$ hyperfine potential with specific spatial form 
using the linked cluster expansion method to eliminate hybrid intermediate states. 

Because
contributions from gluonic excitations and hybrid states are difficult to calculate, 
this work  studies several different parameterizations of the
following generic transverse hyperfine interaction 
\begin{equation}
V_T = \frac{1}{2} \int d {\bf x} d {\bf y} J^{a}_i ({\bf x}) \hat{U}_{ij}({\bf x},
{\bf y})
J^{a}_{j} ({\bf y}) \ ,
\end{equation}
where the kernel $\hat{U}_{ij}$  has the structure
\begin{equation}
\hat{U}_{ij}({\bf x},{\bf y})
=
\left( \delta_{ij} - \frac{\nabla_i \nabla_j}{\nabla^2} \right)_{\bf x}
 \hat{U}(|{\bf x}-{\bf y}|) \ ,
\label{eqhyp}
\end{equation}
reflecting the transverse gauge condition.
In specifying the potential $\hat{U}$ it
is useful to realize that  
perturbatively
$\hat{U} \rightarrow
 \alpha_s/|{\bf x} - {\bf y}|$ where $\alpha _s = g^2/4 \pi$.
Also,
the form of the one gluon
exchange potential and the nonperturbative mixing with hybrids makes it clear that
the hyperfine kernel should not include a confining term. This is important for
infrared divergent gap equations~\cite{Bicudo3}.   Consistent with these points,
the following four kernels are utilized and numerically compared to  document hyperfine
model sensitivity.

The non-relativistic quark model advocates a regulated contact interaction~\cite{gi}.
Thus  model 1 is a simple square well interaction defined by
\begin{equation}
U_1(p)=\left\{ \begin{array}{ll}  0 &  {\rm for} \ p >\Lambda \\
      -U_h  & {\rm for} \ p <\Lambda \end{array} \right.
 \label{squarewell}
\end{equation}
with strength, $U_h$, and  range, $\Lambda$.

Model 2 is a variation of a pure Coulomb potential (reflecting 
a transverse zero mass gluon exchange) and is 
\begin{equation}
U_2(p) = \left\{\begin{array}{ll} C(p) & {\rm for} \ p >m_g \\ 
   -\frac{C_h}{p^2} & {\rm for} \ p <m_g \ . \end{array} \right.
\label{pureCoulomb}
\end{equation}

Similarly, model 3 incorporates a modified Coulomb potential
corresponding to an ultraviolet Coulomb
tail matched to a constant in the infrared. This potential is
\begin{equation}
U_3(p)= \left\{ \begin{array}{ll} C(p)  & {\rm for} \ p >m_g\\
              -C_h  & {\rm for} \ p <m_g \ . \end{array} \right.
\end{equation}

Finally, model 4 is a Yukawa-type potential  corresponding to
the exchange of a constituent gluon with a dynamical mass. This is given by
\begin{equation}
U_4(p)= \left\{ \begin{array}{ll}  C(p) & {\rm for} \ p>m_g \\ 
            -\frac{C_h}{p^2+m_g^2} &  {\rm for} \ p <m_g \ . \end{array}\right.
\label{yukawa}
\end{equation}

For the latter three models, the Coulomb potential $C(p)$ is  the
same as in Eq.~(\ref{SSpotential}) and the constant, $C_h$, is determined by
matching the high and low momentum regions
at the transition scale $m_g$.
In this analysis several different matching points (e.g. $p = m_g, 2 m_g, 3m_g$) for a given 
transition scale $m_g$ were numerically examined  
but  no qualitative differences were found.

In the following, angular 
integrals are denoted by 
\begin{equation}
 V_n(k,q) \equiv \int_{-1}^1 dx V(\ar{\bf k}-{\bf q}\ar) x^n
\end{equation}
where $x = \hat {\bf k} \cdot \, \hat {{\bf q}}$. The
auxiliary  functions
\begin{equation}
 W({|\bf k} - {\bf q}|) \equiv U(|{\bf k} - {\bf q}|)
\frac{x(k^2+q^2)-qk(1+x^2)}{\ar{\bf k}-{\bf q}\ar^2}
\end{equation}
and
\begin{equation}
Z(|{\bf k} - {\bf q}|) \equiv  U(|{\bf k} - {\bf q}|)\frac{1-x^2}{\ar{\bf
k}-{\bf q}\ar^2}
\end{equation}
are also introduced
which arise from the operator structure of Eq.~(\ref{eqhyp}).

\subsection{Gap equation}

Calculations are most conveniently made in momentum space with constituent quark 
operators.  These are 

\onecolumngrid
\noindent
obtained by a Bogoliubov transformation
(BCS rotation) from the current quark basis to a quasiparticle quark basis 
represented by particle, $B$, and antiparticle, $D$, operators
\begin{equation}
\Psi({\bf x}) = \int \frac{d{\bf k}}{(2\pi)^3} e^{i {\bf k} \, \cd \, {\bf x}}
\sum_{\lambda i} ({\cal U}_{{\bf k} \lambda}  B_{{\bf k}\lambda i} +
{\cal V}_{-{\bf k}\lambda}  D^{\dagger}_{-{\bf k} \lambda i}) \hat {\epsilon}_i 
\end{equation}
where $\hat {\epsilon}_i$ is a color vector with $i=  $  1, 2, 3  and $\lambda$ denotes
helicity. The Dirac spinors are functions of the Bogoliubov gap angle $\phi_k$ and can be
express in terms of the Pauli spinors $\chi_{\lambda}$
\begin{eqnarray} { \cal U}_{{\bf k} \lambda} &=& \frac{1}{\sqrt{2}} \left[
\begin{array}{l}
\sqrt{1+\sin\phi_{k}} \, \, \chi_{\lambda}    \\ \sqrt{1-\sin \phi_k}
\, \bm{\sigma} \cdot \hat{{\bf k}} \, \, \chi_{\lambda}
\end{array}
\right] \\
%
{\cal  V}_{{\bf -k} \lambda}&=& \frac{1}{\sqrt{2}}\left[ \begin{array}{l}
-\sqrt{1-\sin \phi_k} \, \, \bm{\sigma}\cdot \hat{\bf k}
\, \, i\sigma_2\chi_{\lambda} \\
\sqrt{1+\sin \phi_k} \, \, i\sigma_2\chi_{\lambda}\end{array} \right] \ .
\end{eqnarray}
Note the additional factor $i\sigma_2$ in the  spinor $\cal V_{{\bf -k}
\lambda}$ which
differs from the convention used in Refs.~\cite{LC1,LC2}.
The Bogoliubov angle can be related to a running quark mass, $M_q(k)$, and energy,
$E(k) = \sqrt {M^2_q(k) + k^2}$, by
\begin{eqnarray}
s_k &\equiv& \sin \phi_k = \frac{M_q(k)}{E(k)} \label{mck} \\
 c_k &\equiv& \cos \phi_k = \frac{k} {E(k)} \ .
\end{eqnarray}
At high $k$, $M_q(k) \rightarrow m_q$, while for low $k$  a constituent quark mass
can be extracted, $\mathcal{M}_q = M_q(0)$.

Minimizing the vacuum energy with respect to the gap angle
yields the mass gap equation
\begin{eqnarray}
k \ s_k - m_q \ c_k = \int \frac{d{\bf q}}{12\pi^3}
\left[ (s_k c_q x - s_q c_k)
V(\ar {\bf k} - {\bf q} \ar) 
-2 c_k s_q U(\ar {\bf k} - {\bf q}\ar)  + 2c_q s_k W(\ar {\bf k} - {\bf
q}\ar)  \right]
\end{eqnarray}
which, after angular integration, reduces to
\begin{eqnarray}
k \ s_k - m_q \ c_k =  \int_0^\infty \frac{q^2dq}{6\pi^2}
\left[   s_k c_q (V_1 + 2 W_0) -  s_q c_k (V_0 + 2 U_0) \right] \ .
 \label{gapEq}
\end{eqnarray}
Finally, the quasiparticle self-energy  is
\begin{eqnarray}
\epsilon_k= m_q \ s_k + k \ c_k - \int_0^\infty
\frac{q^2 dq}{6\pi^2} \left[s_k s_q (V_0 
+ 2 U_0)
+c_k c_q (V_1 +2 W_0)
\right] \ .
\label{onebody}
\end{eqnarray}

\subsection{Meson RPA equations}

The pion RPA creator operator, $\mathcal{X}_{\alpha \beta} B^\da_\alpha
D^\da_\beta - \mathcal{Y}_{\alpha \beta} B_\alpha D_\beta $, contains two
wavefunctions with color $a$, $b$, spin
$\alpha$, $\beta$,
momentum,  ${\bf k}$, and radial, $\nu$, indices  given by
\begin{eqnarray}
\mathcal{X}^\nu({\bf k}) &=& \frac{1}{\sqrt{4 \pi}} \frac{i
(\sigma_2)_{\alpha \beta}}{\sqrt{2}} \frac{\delta_{ab}}{\sqrt{3}}
X^\nu (k)
\\ \nonumber
\mathcal{Y}^\nu ({\bf k}) &=& \frac{1}{\sqrt{4 \pi}} \frac{i
(\sigma_2)_{\alpha \beta}}{\sqrt{2}} \frac{\delta_{ab}}{\sqrt{3}}
Y^\nu (k) \ .
\end{eqnarray}
Diagonalizing  the effective Hamiltonian  in the RPA
representation yields two coupled radial  equations
valid for any equal quark mass  pseudoscalar meson
\begin{eqnarray} \label{pionRPA}
 2 {\epsilon}_k \  X^\nu (k)+
\int_0^\infty
\frac{q^2dq}{6\pi^2} \left[  K(k,q)  X^\nu (q)
 +  K^{\prime} (k,q) { Y}^\nu (q) \right] 
&=& M_{\pi}^{\nu} { X}^\nu(k) \\ \nonumber
2 {\epsilon}_k \ { Y}^\nu(k)+
\int_0^\infty
\frac{q^2dq}{6\pi^2}\left[  K(k,q) { Y}^\nu(q) +  K^{\prime} (k,q) { X}^\nu (q)
\right] &=& -M_{\pi}^\nu { Y}^\nu(k) \ ,
\end{eqnarray}
with kernels
\begin{eqnarray} \label{kernel}
K(k,q)&=& (1+s_k s_q)V_0 + 2(1-s_ks_q)U_0
+c_k c_q (V_1 - 2W_0)
\\ 
K^{\prime}(k,q)&=&  (1-s_ks_q)V_0 + 2(1+s_k s_q)U_0 
- c_k c_q(V_1 -2W_0) \ . \label{kernelp}
\end{eqnarray}

\newpage

The above equations yield a zero mass pion in the chiral limit~\cite{Bicudo1,Scadron,mrt98} as
demonstrated by  combining the gap and  self energy
equations to obtain
$$
s_k \epsilon_k = m_q -  \int \frac{q^2dq}{6\pi^2}
s_q(V_0 +2 U_0)  \ ,
$$
and then substituting this expression into
the RPA equations,  first multiplied by $s_k$. For $m_q =0$, 
${\mathcal X}(k)$ and ${\mathcal Y}(k)$ become proportional to $s_k$. This immediately yields the
eigenvalue  $M_{\pi}=0$  in accord with Goldstone's theorem.
We have numerically confirmed this to a precision of 5 keV in solving 
the RPA equations.


Constructing the $\rho$ and other vector meson RPA wavefunctions requires 
three spin projections. Also, both
$S$ and $D$ orbital waves now contribute and the general solution is
\begin{eqnarray}
\bm{\mathcal{X}}^\nu&=&\bm{\mathcal{X}}_s^\nu+\bm{\mathcal{X}}_d^\nu \\
\bm{\mathcal{Y}}^\nu&=&\bm{\mathcal{Y}}_s^\nu+\bm{\mathcal{Y}}_d^\nu \ .
\end{eqnarray}
It is  reasonable  to assume exact isospin symmetry (degenerate quark
masses, $m_u = m_d$), which permits suppression of this quantum number. The four
wavefunctions having unit norm are then
\begin{eqnarray}
\mbox{\boldmath   $\mathcal{X} $\unboldmath}_s^\nu &=&  \frac{1}{\sqrt{4\pi}}
\frac{ \mbox{\boldmath   $\mathcal{\sigma} $\unboldmath}
i\sigma_2}{\sqrt{2}} \frac{\delta_{ab}}{\sqrt{3}} X_s^\nu(k)
\\ \nonumber
\mbox{\boldmath   $\mathcal{X} $\unboldmath}_d^\nu &=&  \frac{1}{\sqrt{4\pi}}
\frac{3}{2}(\hat{\bf k}\cd \mbox{\boldmath   $\mathcal{\sigma}
$\unboldmath}\, \hat{\bf k} -\frac{1}{3}
\mbox{\boldmath   $\mathcal{\sigma} $\unboldmath}) i \sigma_2
\frac{\delta_{ab}}{\sqrt{3}} X_d^\nu(k)
\\ \nonumber
\mbox{\boldmath   $\mathcal{Y} $\unboldmath}_s^\nu &=&  -\frac{1}{\sqrt{4\pi}}
\frac{\mbox{\boldmath   $\mathcal{\sigma} 
$\unboldmath}i\sigma_2}{\sqrt{2}}
\frac{\delta_{ab}}{\sqrt{3}} Y_s^\nu(k)
\\ \nonumber
\mbox{\boldmath   $\mathcal{Y} $\unboldmath}_d^\nu &=&  -\frac{1}{\sqrt{4\pi}}
\frac{3}{2}
(\hat{\bf k}\cd \mbox{\boldmath   $\mathcal{\sigma}
$\unboldmath}\, \hat{\bf k} -\frac{1}{3}
\mbox{\boldmath   $\mathcal{\sigma} $\unboldmath})
i\sigma_2 \frac{\delta_{ab}}{\sqrt{3}} Y_d^\nu(k) \ .
\end{eqnarray}

Exploiting the symmetry of the RPA kernels,
under transposition and simultaneous ${\bf k}$ $\leftrightarrow$ ${\bf q}$
exchange, reduces
the number of independent kernels to six:
$K_{XX}^{ss}$,  $K_{XX}^{sd}$, $K_{XX}^{dd}$,
$K_{XY}^{ss}$, $K_{XY}^{sd}$, $K_{XY}^{dd}$. The ten other required
kernels can be obtained from these using
$K_{YY}=K_{XX}$, $K_{XY}=K_{YX}$ for all four angular momentum
combinations and $K^{ds}(k,q)=K^{sd}(q,k)$ for all four $X$-$Y$ combinations.
The $\rho$ RPA equations are 
\begin{equation}
 2 \epsilon_k \left( \begin{array}{c} X^\nu_s \\ X^\nu_d \\ Y^\nu_s \\ Y^\nu_d
\end{array} \right) +
\int_0^\infty \frac{q^2dq}{6\pi^2} \left[ \begin{array}{cccc}
K_{XX}^{ss} & K_{XX}^{sd} & K_{XY}^{ss} & K_{XY}^{sd} \\
K_{XX}^{ds} & K_{XX}^{dd} & K_{XY}^{ds} & K_{XY}^{dd} \\
K_{YX}^{ss} & K_{YX}^{sd} & K_{YY}^{ss} & K_{YY}^{sd} \\
K_{YX}^{ds} & K_{YX}^{dd} & K_{YY}^{ds} & K_{YY}^{dd}
\end{array}\right]
\left( \begin{array}{c} X^\nu_s \\ X^\nu_d \\ Y^\nu_s \\ Y^\nu_d
\end{array} \right)
= M^\nu_\rho \ \
\left( \begin{array}{c} X^\nu_s \\ X^\nu_d \\ -Y^\nu_s \\ -Y^\nu_d
\end{array}\right)
\ , \end{equation}
and the integration is performed 
after multiplication with the column wavefunction  vector.

Finally, the six independent kernels are
\begin{eqnarray}
K_{XX}^{ss}&=& \frac{1}{6}\left[3(1+s_k)(1+s_q) V_0 +
(1-s_q)(1-s_k)(4 V_2-V_0) +2 c_k c_q(3V_1  +
2 U_{1} + 2 k q Z_0) \ + \nonumber \right. \\  && { } \left.
   2 (1+s_k)(1-s_q) (-U_{0}
+2 k^2 Z_0) +
2 (1-s_k)(1+s_q)(-U_{0} +2 q^2 Z_0 ) \right]
\\
K_{XX}^{sd}&=& \frac{\sqrt{2}}{6} \left[
(1-s_q)(1-s_k)(V_2-V_0)
 +
(1+s_k)(1-s_q)(-2 U_{0}
+k^2 Z_0)+ 2
c_kc_q(2 U_{1}-kq Z_0) \ + \nonumber \right. \\  && { } \left.
(1-s_k)(1+s_q)(U_{0}-3 U_{2} +q^2 Z_0) \right]
\\
K_{XY}^{ss}&=& \frac{1}{3}\left[ -(1-s_ks_q)
V_0
-   (1+s_k)(1+s_q)U_{0}
+ (1-s_k)(1-s_q)(U_{0} - 2 U_{2})
+ c_k c_q (V_1 - 2U_{1} + 2kq
Z_0)\right]
\\
K_{XY}^{sd}&=& \frac{\sqrt{2}}{6} \left[ 2(1+s_k)(1-s_q)V_0 +
(1+s_q)(1-s_k)(3 V_2- V_0)  +
(1+s_q)(1+s_k)
(-2U_{0}+3 q^2 Z_0)  \nonumber \right. \\  && { } \left.
  + \
(1-s_k)(1-s_q)(- U_{0}- U_{2} +3
k^2 Z_0)-2 c_kc_q (2 V_1 + 2 U_{1}+kq Z_0) \right]
\\
K_{XX}^{dd}&=& \frac{1}{12} \left[
3(1+s_k)(1+s_q)(3V_2-V_0) + (1-s_k)(1-s_q)(V_2+5V_0)
+ 4c_kc_q (3V_1 +4 U_{1} +k q Z_0) \ + \nonumber \right. \\  && { } \left.
  (1+s_k)(1-s_q)
(-5 U_{0}-3 U_{2} \ +
 (k^2
+9q^2) Z_0)+(1-s_k)(1+s_q)(-5 U_{0}-3
U_{2}
+(9k^2+ q^2) Z_0)  \right]
\\
K_{XY}^{dd}&=&\frac{1}{12} \left[ 2(1-s_k s_q)
(3 V_2- V_0)-4c_kc_q (V_1 + 4   U_{1} - k q
Z_0)
+
(1+s_k)(1+s_q)(U_{0}-9 U_{2} +
 3 (k^2+q^2) Z_0) \right. \nonumber \\  && {} \left.
+ \ (1-s_k)(1-s_q)( -7 U_{0}- U_{2}
+3 (k^2+q^2) Z_0 )  \right] \ .
\end{eqnarray}

\newpage
\twocolumngrid

\section{Numerical results and discussion}
\subsection{Meson spectra}

The numerical techniques for solving the gap equation and diagonalization for
the meson eigenvalues are given in Refs.~\cite{ls2,LC1,LC2,ssrenorm}.
The four hyperfine model interactions were each determined by fitting
the  charmonium $\eta_c$-$J/\Psi$ splitting.  In addition to adjusting the
potential parameters, $U_h$, $\Lambda$ for model 1 and $m_g$ for models
2, 3 and 4, it was also necessary to
significantly reduce the current charm quark mass from values typically used 
which is discussed  below.
The  hyperfine potentials  are summarized in Table
\ref{parameters}.

\begin{table}[h]
\caption{\label{parameters}
Different hyperfine effective interactions.}
\begin{ruledtabular}
\begin{tabular}{lllr}
 Model       & Parameters  & &\\ \hline
$1. \  \rm {square \  well}$ &$ \rm {U_h} =	5$  GeV$^{-2}$,	& $\Lambda
= 3  m_g = 1.95$ GeV    	&\\
$2. \ \rm {Coulomb}$
                & $m_g = 0.43$ \ GeV  &  &\\
$3. \ \rm {modified \  Coulomb}$         &  $m_g = 0.6$  GeV &  	&\\
$4. \ \rm {Yukawa}$ & $m_g = 0.6$  GeV        &    	&\\
\end{tabular}
\end{ruledtabular}
\end{table}

Using these potential parameters the remaining light and heavy
pseudoscalar and vector meson spectra  were then predicted.
The $\pi, \rho, \eta_c$, $J/\Psi$, $\eta_b$ and $\Upsilon$ ground and excited
states are listed in Table \ref{spectrum}.
While all four models
provide similar, reasonable
meson descriptions, potential 4 emerges as the preferred model.  Note that it was again
necessary to reduce the current quark masses.  

\begin{table}[h]
\caption{\label{spectrum}
Calculated masses, condensates and data in MeV (rounded to the 
nearest 5 MeV).}
\begin{ruledtabular}
\begin{tabular}{c|ccccc}
       Quantity & Model 1 & Model 2 &Model 3 & Model 4& experiment \\ 
\hline
$^{\dagger}m_{u}=m_{d}$ &	1 	& 1      	  &  1   & 1  & 1.5
- 8.5 \\
$^{\dagger}m_{c}$ &	640 	& 520     &  530 & 510  & 1000 - 1400 \\
$^{\dagger}m_{b}$ &     3330     & 2800     & 2750   & 2710 & 4000-4500\\
$\mathcal{M}_{u}=\mathcal{M}_d$ & 85      & 145  &  85 & 100 &200 - 300 \\
$\mathcal{M}_{c}$ &	1090	& 1130    &  1090& 1090  & 1500 \\
$\mathcal{M}_{b}$ &     4025    & 4020    & 3980 & 3965  & 4600-5100 \\
$-<\bar{q}q>^{\frac {1} {3}}$  &  150     & 200 &  165 & 180  & 220-260 \\
$M_{\pi}$ &	195	& 150      	  &  190 & 190  & 138 \\
$M_{\pi(1300)}$ &	1430& 1150        & 1350 & 1370  & 1300\\
$M_{\pi(1800)}$ &	2170& 1650        & 2085 & 2100  & 1801\\
$M_{\rho}$ &	820  & 755      	  &  780 & 795  & 771\\
$M_{\rho(1450)}$ &	1480& 1150        & 1405 & 1420  & 1465\\
$M_{\rho(1700)}$ &	1725& 1305     	  & 1605 & 1620 & 1700\\
$M_{\eta_c(1S)}$ &      2980& 2950        & 2990 & 2985 & 2980\\
$M_{\eta_c(2S)}$ &	3660& 3400    	  & 3615 & 3625 &  \footnote{The
Belle collaboration~\cite{Belle} reports $3654 \pm 0.014$}3631\\
$M_{\eta_c(3S)}$ &	4210& 3720     	  & 4090 & 4100 & ?\\
$M_{J/\Psi(1S)}$ &	3110& 3130        & 3110 & 3130 & 3097\\
$M_{\Psi(2S)}$ &	3740& 3470    	  & 3670 & 3680 & 3686\\
$M_{\Psi(3770)}$ &	3780& 3490     	  & 3685 & 3695 & 3770\\
$M_{\eta_b}$ &      9395   &  9360       & 9415 & 9395  & ?\\
$M_{\Upsilon}$ &    9465   &  9440       & 9470 & 9460 & 9460\\
$M_{\Upsilon(2s)}$ &  9915 &  9705       & 9880 & 9870  & 10023\\
\end{tabular}
\end{ruledtabular}
$^{\dagger} \text {adjusted}$
\end{table}

\bigskip

\begin{figure}[h]
\includegraphics[width=8cm]{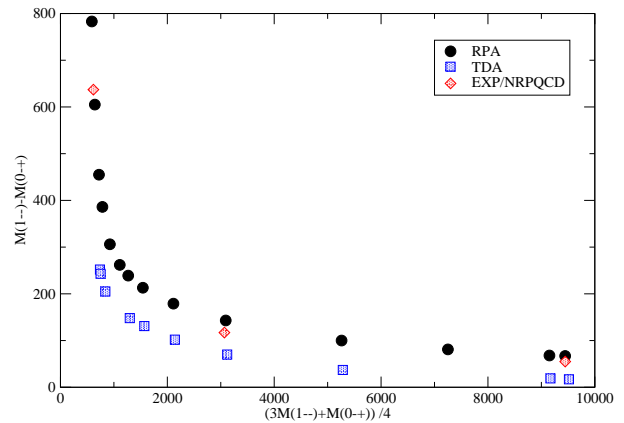}
\caption{\label{figsplit}Hyperfine splitting versus spin averaged meson
multiplet mass.}
\end{figure}

It is  significant that the RPA Hamiltonian approach, using any of the four
hyperfine interactions, can
simultaneously describe both the large $\pi$-$\rho$ mass difference and the small
$\pi'$-$\rho'$ and charmonium splittings.  Figure~\ref{figsplit} further illustrates this by 
comparing the RPA (solid circles), TDA (squares) and observed (diamonds)
hyperfine splittings versus the spin averaged pseudoscalar and vector meson mass.
Notice that similar to observation both the RPA and TDA yield a rapid decline in
the  spin  splittings with increasing meson mass, however
only the RPA can describe the sizeable $\pi$-$\rho$ difference.  This
is because the RPA consistently implements chiral symmetry while the non-chiral TDA
predicts a pion mass that is too large (about 500 MeV).
Similar to the findings of Ref.~\cite{LC1}, chiral symmetry is clearly the dominant effect in 
the large $\pi$-$\rho$ splitting, accounting for almost 70\% (roughly 400
MeV) of the mass difference. 

It is insightful to contrast this with 
non-relativistic quark model treatments which use a constituent quark mass about half the
multiplet average and describe the splitting as a $1/\mathcal{M}^2$
dependence characteristic
of relativistic corrections.  Because the quarkonium Bohr orbit scales inversely
with  quark mass, it is possible for these models to describe both light and heavy meson
splittings with the same short ranged hyperfine potential although it does require tuning of
parameters.  Indeed as Ref.~\cite{gi} details, a comprehensive meson
description can be obtained by using a potential with a complicated, mass dependent
short ranged smearing. Alternatively, the RPA-BCS formalism, which dynamically generates a
constituent mass, 
reproduces this behavior via chiral symmetry and a simpler, weaker hyperfine interaction. The TDA,
which also incorporates the same quasiparticle/constituent running masses as the RPA, in general
provides a qualitatively comparable   spin splitting description except for the
$\pi$-$\rho$ difference.  Of course by increasing the hyperfine strength it would be
possible for the TDA to account for the large
$\pi$-$\rho$ splitting, however, this enhanced interaction would then generate an  
overprediction for the
$\pi'$-$\rho'$ and other splittings. For a comprehensive description the TDA would most likely
require a more complicated hyperfine interaction with tuning 
and in this
sense shares the same difficulties as constituent, non-chiral models mentioned in the
introduction.    The attractive feature of the RPA-BCS approach is the ability to obtain a
good description with minimal parameters which is a common goal in all approaches
to QCD and hadron structure.

Application to the isoscalar hyperfine splitting is not possible
without a proper $\eta$-$\eta^\prime$ mixing calculation.
However, using the strange current mass of 25 MeV and the 
preferred potential, model 4, the splitting for the pure 
$s\overline{s}$ mesons is 300 MeV, corresponding to a pseudoscalar mass of 
720 MeV and a $\phi$ meson at 1020 MeV.
The extracted strange constituent mass was 192 MeV. 
As the physical $\eta$ and $\eta^\prime$ masses are respectively 547 and
958 MeV,
mixing effects are significant and it will be of interest to see the importance
of the hyperfine interaction in a rigorous mixing analysis.

In addition to an improved meson spin spectrum, three of the
model hyperfine interactions markedly increase the quark condensate which
previous analyses~\cite{orsay,LC1,LC2,ls2} predicted  too low, around
$\langle \bar q q \rangle = - (110\  {\rm MeV})^3$. 
This was also noted in Ref.~\cite{aspk}.
For models 2, 3 and 4 the new condensate varies
between $-(165\ {\rm MeV})^3$ and $-(200\ {\rm MeV})^3$ in the chiral limit, a noticeable 
improvement but still below accepted values spanning the interval between $-(220\ {\rm
MeV})^3$ and $-(260\ {\rm
MeV})^3$. The pion decay constant, $f_\pi$, also improves but only marginally. 
These shifts are in the correct direction,
as  first noted  by Alkofer and Laga\"e~\cite{Lagae}, but complete agreement in this model
is not possible without generating very large
self energies which distort the meson spectrum.  Because the decay constant is
a matrix element connecting the ground state (model vacuum), the low calculated values
also reflect shortcomings with the BCS vacuum.  As detailed in Ref.~\cite{LC2}
the use of the superior RPA vacuum significantly increases $f_\pi$ (although
not quite to the physical value). It would therefore be interesting 
to repeat this hyperfine calculation with an improved vacuum. 


Naively,  it is expected that the transverse potential should, as in
the quantum mechanical quark model, decrease the mass of  pseudoscalar
states.  However, it is important to distinguish between level splitting and
absolute level shifts.
The hyperfine interaction does indeed provide a level splitting with difference
proportional to hyperfine strength, but it also increases the
quasiparticle
self-energy and thus the
effective constituent quark mass as well.
Consequently, both  pseudoscalar and
vector meson
masses increase which then in turn requires a  reduction in the current quark mass
to reproduce the observed spectra. 

This reduction in current quark mass was also necessary to describe states in 
bottomonium. The reported but unconfirmed $\eta_b$ state has a mass of 9300 
MeV~\cite{RPP}, clearly below predictions (see Table \ref{spectrum}) which
are closer to non-relativistic perturbative QCD  (NRPQCD) and 
lattice calculations~\cite{pqcd} that predict a much smaller 
$\Upsilon$-$\eta_b$
splitting of 
about 40 MeV with an error of about 20 to 30 MeV.
A recent paper~\cite{bak} lowers this error to 10 (th$)^{+9}_{-8}$ ($\delta \alpha_s$) MeV 
reflecting
uncertainties in theory and  the strong coupling constant.
The NRPQCD 
calculations, in particular, suffer from uncertainty in nonperturbative 
corrections that are usually parameterized in terms of condensates. Hence 
model calculations are still needed and the RPA
predicts a splitting in rough agreement, but somewhat larger, than 
these theoretical expectations.  Interestingly, the structure of the coupling (see Eqs.
(\ref{pionRPA}), (\ref{kernel}) and (\ref{kernelp})) is such  to increase  the splitting from the
TDA value of around 20 MeV to the RPA prediction of 60 MeV. Although this is a minimal, secondary
effect  when compared to the 
absolute mass scale involved, it is still important for the relatively 
small hyperfine separation. 
This analysis therefore predicts that the 
$\eta_b$ meson mass should be around 9400 MeV and  
new results from spectroscopic studies at the B-factories are eagerly awaited. 

\subsection{Comparison to other hadronic approaches}

\begin{figure}[t]
\includegraphics[width=7.5cm,angle=-90]{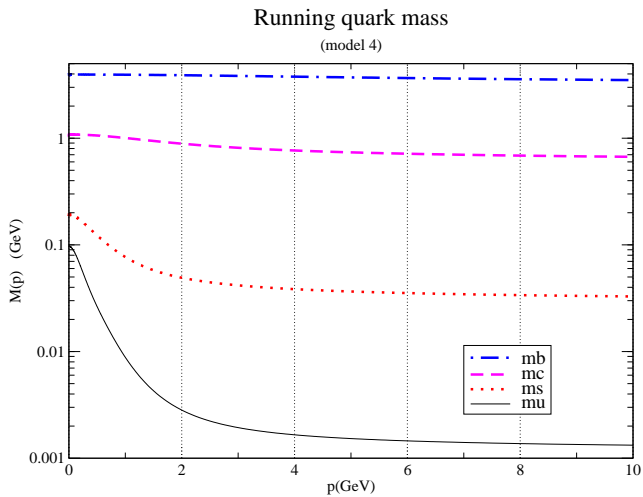}
\caption{\label{runmfig} Constituent $u/d$, $s$, $c$ and $b$ running quark masses from
Eq. (\ref{mck}).}
\end{figure}

It is instructive to make contact with alternative formulations and
to further discuss 
the smaller current and constituent quark masses in our model.  It is well known
that a running current quark mass emerges    in one loop QCD and values can be extracted at the
perturbative 
$M_Z$ scale.
Usually quoted, however, are their values at a much lower renormalization point 
such as 2  GeV in the $\overline{MS}$ scheme. While the "experimental" bare quark values listed 
in Table \ref{spectrum} are obtained from measurement, they rely on
significant model input either from
chiral  perturbation theory, as in the case of the $u$, $d$, $s$ masses, or from
heavy quark effective theory for the $c$, $b$ quarks.
They are thus not directly observable and entail both uncertainty and also
ambiguity~\cite{creutz}.  It is therefore more appropriate
to regard them as  parameters in the Lagrangian  (Hamiltonian), subject to
renormalization.  
Since our model  
kernel has been previously renormalized the only remnant
of current quark mass running is the momentum dependence of the dressed
quark mass $M_q(p)$.  Effectively, the current quark renormalization point
dependence has been converted into a constituent mass momentum dependence
analogous to the Schwinger-Dyson treatment.  Therefore when comparing to current masses
from other approaches one should use the constituent running mass
evaluated at the alternative model's current mass scale, e.g. 
$M_q (p = 2$ GeV), and not the smaller $m_q$ parameter appearing in our 
Hamiltonian. Figure \ref{runmfig} plots 
the running dependence of our model 4 dressed quark masses for different flavors.  
As this figure
indicates at $p$ = 2 GeV,  the scaled effective current quark masses are still lower than in 
other approaches, however they are much larger than our bare Hamiltonian values 
and provide a more realistic comparison.  Related, the larger constituent
masses in conventional quark models (ranges listed in Table II under "experiment") partially
reflects missing dynamics from field theoretical self-energies that we explicitly include.
The validity of any quark approach should consequently be judged more by the
robustness  of observable prediction (e.g. spectrum) rather than specific quark values.
This is the case in the present analysis
as indicated in Table
\ref{spectrum}, where the resulting quark masses are small
when compared  to the typically quoted  values but the predicted meson masses are reasonable. 
We submit these smaller current quark values are representative of this
simple, minimal parameter approach because the
masses were uniformly small for all four, markedly different hyperfine interactions.
An improved, rigorous treatment entailing a complicated combined quark-gluon
sector diagonalization of the exact hyperfine Hamiltonian, Eq. (\ref{Hqg}), is
in progress which will firmly ascertain if small quark values are required.

Finally,  predictions for the ``J parameter'' 
\begin{equation}
J = M_R {d M_V \over d M_{Ps}^2}
\end{equation}
are presented which has been proposed as a reliability measure for quenched lattice
computations of light hadron masses~\cite{J}. Here $M_{Ps}$, $M_V$ are the pseudoscalar, vector
meson masses and
$M_R$ is the (reference) vector mass determined by the intersection of the line $M_V = 1.8 \
M_{Ps}$ with the plot of $M_{Ps}^2$ versus $M_V$.
If the  vector meson
mass is linear in the current quark mass and, as indicated by Eq.~(\ref{gmor}), the
pseudoscalar scales as the square root, then a sensitive lattice chiral extrapolation 
is not required to evaluate $J$ and the attending errors can be avoided.
Results from unextrapolated quenched
lattice simulations~\cite{J} predicts $J =  0.37$ which should be contrasted with
the estimate $J = 0.48$ using the physical
$\pi$, $\rho$, $K$ and $K^*$ masses.  The difference
reflects the need to include dynamical
fermions in the lattice calculations. 

\begin{figure}[h]
\includegraphics[width=5cm,angle=-90]{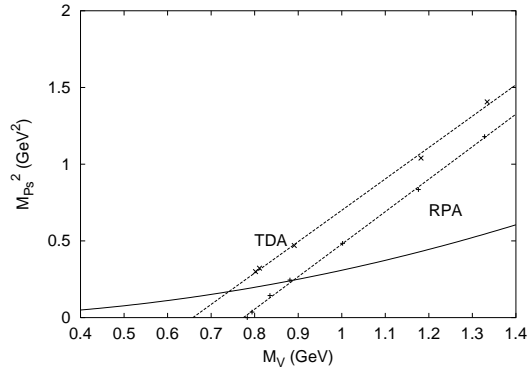}
\caption{\label{Jfig} Determining the $J$ parameter. The solid curve corresponds to
$M_V = 1.8 \ M_{Ps}$ while the other lines are linear fits to the TDA and RPA predictions.}
\end{figure}


Figure \ref{Jfig} shows the TDA and RPA masses and the 
curve $M_V = 1.8 \ M_{Ps}$.  As
anticipated, the RPA points scale
linearly and extrapolate to zero pion mass at a 
vector mass of approximately 780 MeV. A linear fit to the RPA gives a reference
mass of $M_R = 880$ MeV and $J = 0.42$, in more reasonable agreement with the estimate from 
data. Surprisingly, the TDA points also scale linearly even though they do not yield a
zero mass pion in the chiral limit. This lowers the reference vector mass and produces a
$J$ parameter of $0.36$. 
It will prove instructive to confront these predictions with dynamical quark
lattice simulations especially those using  realistic sea quark masses.

\section{Conclusion}

As a consequence of this study, the relative importance of chiral symmetry
and the hyperfine interaction is clearer: spin splittings in heavy quark systems are not 
governed by chiral symmetry and only require a 
hyperfine interaction. However,  for light
mesons chiral symmetry is important and is essential for describing
the $\pi$-$\rho$ mass difference in a minimally parameterized, heavily constrained model
such as the one advocated here.  Indeed the RPA-BCS many-body
approach provides a reasonable description of the pseudoscalar-vector spectrum for
both light and heavy mesons with a common Hamiltonian containing only the
current quark masses and two dynamical parameters.  By explicitly incorporating
this important symmetry of QCD, a small pion mass is dynamically generated without the
necessity of tuning a complicated hyperfine potential as typically done in conventional
quark models. Furthermore, including a hyperfine interaction in this many-body approach 
improves both the pion decay constant and the quark condensate predictions which previously have
been calculated too low. 
The hyperfine interaction also enhances the self energy contribution
to the quark kinetic energy which necessitates using much smaller current
quark masses. 
Lastly, 
the RPA $J$ parameter is closer to data than
quenched lattice results and it will be interesting to  compare 
with dynamical quark lattice simulations.

Future work includes reanalyzing the glueball, meson and hybrid  spectra 
with the hyperfine potential
and examining other short range 
interactions such as the tensor, $\alpha_i \alpha_j$ - $\alpha_i \alpha_j$,
and higher
dimensional terms from excluded Fock space components~\cite{Bicudo3}.
Extensions of this approach to highly excited hadron states will also be of interest, based 
upon the need~\cite{ess2} for new   relativistic, chirally invariant models with
a nontrivial vacuum.
Finally, investigations of baryons should also be fruitful as previous non-hyperfine
calculations~\cite{L2} only predict about half of the
observed $N$-$\Delta$ splitting.

\acknowledgements

F.  Llanes and S. Cotanch thank P. Bicudo and E. Ribeiro for useful comments. S. Cotanch
also acknowledges T. Hare for effective advice.  E.
Swanson  is grateful to R. Woloshyn for a helpful observation. This work
was supported by Spanish grants FPA 2000-0956, BFM 2002-01003 (F.L-E.) and
the Department of Energy grants DE-FG02-97ER41048 (S.C.), DE-FG02-87ER40365
(A.S.)
and DE-FG02-00ER41135, DE-AC05-84ER40150 (E.S.).


\end{document}